\RequirePackage{fix-cm}
\documentclass[smallextended,refree]{svjour3}       
\usepackage[utf8]{inputenc}
\smartqed  
\usepackage{amssymb}
\usepackage{latexsym}
\usepackage{graphicx}
\begin{document}

\title{
End-point singularity of the 
$XY$-paramagnetic phase boundary
for the 
$(2+1)$D $S=1$
square-lattice $J_1$-$J_2$
$XY$ model
with the single-ion anisotropy
}
\subtitle{}


\author{Yoshihiro Nishiyama
}

\institute{Department of Physics, Faculty of Science,
Okayama University, Okayama 700-8530, Japan }

\date{Received: date / Accepted: date}

\maketitle

\begin{abstract}

The two-dimensional quantum spin-$S=1$ square-lattice $J_1$-$J_2$ $XY$ model
with the single-ion anisotropy $D$
was investigated numerically, placing an emphasis
on the end-point singularity of the phase boundary
separating the $XY$ and paramagnetic phases
in proximity to
the fully frustrated point, $J_2/J_1 \to 0.5^{-}$.
We employed the exact diagonalization method
to circumvent the negative sign problem of the quantum Monte Carlo method,
and evaluated the fidelity susceptibility $\chi_F$
as a probe to detect the phase transition.
As a preliminary survey, for
an intermediate value of $J_2/J_1$,
the $D$-driven $XY$-paramagnetic phase transition 
was investigated via the probe $\chi_F$.
It turned out 
that the criticality belongs
to
the $3$D-$XY$ universality class.
Thereby, 
the $\chi_F$ data were cast into the crossover-scaling formula
with the properly scaled
distance from the multi-critical point, $0.5-J_2/J_1$.
The set of multi-critical indices were obtained,
and compared to 
those of the quantum Lifshitz criticality.

\end{abstract}

\section{\label{section1}Introduction}

The quantum spin-$S=1/2$ triangular-lattice $XXZ$ model
\cite{Wessel05,Heidarian05,Gan09,Hassan07,Boninsegni05}
has been investigated
as a realization of the super-solid phase
\cite{Boninsegni12},
relying on
the equivalence between 
the hard-core-boson and quantum-spin operators \cite{Matsubara56,Roscilde07}.
Intriguingly,
the super-solid phase is stabilized
by the 
inter-site Coulomb repulsions.
Here, a key ingredient is that this model
is free from the
negative sign problem of the quantum Monte Carlo (QMC) method,
and the large-scale QMC data are available.
Meanwhile, as for the 
hard-core ($S=1/2$) \cite{Chen17,Tu20} 
and
soft-core \cite{Dong17} boson models
on the
{\em square} lattice,
it has been found that 
the ``kinetic frustration'' \cite{Chen17}
(namely, the frustrated $XY$ interaction)
stabilizes 
a novel type of super-solid phase, 
the so-called half super-solid phase.
The kinetic frustration 
renders the negative sign problem as to the QMC simulation.
Hence,
to cope with this difficulty,
there have been employed
a variety of theoretical techniques,
such as 
the cluster mean-field theory \cite{Chen17},
infinite projected entangled-pair state \cite{Tu20},
exact diagonalization \cite{Tu20},
and tensor network state \cite{Dong17}
methods.

In this paper,
we investigate the $S=1$ square-lattice $J_1$-$J_2$
$XY$ model with the single-ion anisotropy
by means of the exact diagonalization method.
We dwell on the end-point singularity (multi-criticality)
of the phase boundary separating the $XY$ and paramagnetic phases
at the fully frustrated point $J_2/J_1 \to 0.5^-$.
As a probe to detect the phase transition
\cite{Quan06,Zanardi06,HQZhou08,Yu09,You11},
we utilize the fidelity susceptibility \cite{Uhlmann76,Jozsa94,Peres84,Gorin06},
which is readily accessible via the exact diagonalization scheme.
Because 
the scaling dimension of the fidelity susceptibility is 
larger than that of specific heat \cite{Albuquerque10},
the
underlying (multi-)criticality
is captured
rather sensitively via the fidelity susceptibility;
note that the specific-heat critical exponent 
takes a negative value
\cite{Campostrini06,Burovski06}
for the 3D-$XY$ universality class \cite{Roscilde07}.

To be specific, we present the Hamiltonian
for the $S=1$ square-lattice $J_1$-$J_2$
$XY$ model with the single-ion anisotropy
\begin{equation}
\label{Hamiltonian}
{\cal H}= -J_1  \sum_{\langle ij \rangle} (S_i^xS_j^x+S_i^yS_j^y)
+J_2  \sum_{\langle\langle ij \rangle\rangle} (S^x_iS^x_j+S^y_iS^y_j)
+D \sum_{i=1}^N (S^z_i)^2
   .
\end{equation}
Here, the quantum spin $S=1$ operator ${\mathbf S}_i$
is placed at each square-lattice point $i=1,2,\dots,N$.
The summations,
$\sum_{\langle ij \rangle}$ and 
$\sum_{\langle \langle ij \rangle\rangle}$,
run over all possible nearest-neighbor
$\langle ij \rangle$
and
next-nearest-neighbor 
$\langle \langle ij\rangle\rangle$
pairs, respectively.
The parameters, $J_1$ and $J_2$, are
the respective coupling constants.
Hereafter,
the former is regarded as the unit of energy $J_1=1$.
The latter induces the magnetic frustration.
The symbol $D$ denotes the single-ion anisotropy.
In the boson language \cite{Matsubara56,Roscilde07},
the $J_{1,2}$ terms are the ``kinetic frustration'' \cite{Chen17},
and the $D$ term is interpreted as the on-site Coulomb repulsion
among the soft-core bosons.

\begin{figure}
\includegraphics{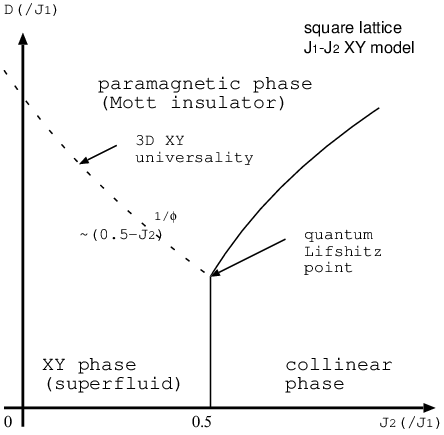}
\caption{ \label{figure1}
A schematic ground-state phase diagram 
\cite{Dutta98}
is presented
for the $S=1$
square-lattice $J_1$-$J_2$ $XY$ model
with the single-ion anisotropy $D$ 
(\ref{Hamiltonian}).
The limiting cases of  $D=0$ \cite{Bishop08}
and $J_2=0$
\cite{Pires11}
have been investigated in depth.
For large $D$, the paramagnetic phase is realized.
In the small-$D$ regime, the
$XY$ and 
	collinear
	phases appear, as the second-neighbor interaction $J_2$ changes.
The dashed (solid) line stands for the continuous (discontinuous) phase transition
\cite{Dutta98}.
The $XY$-paramagnetic phase boundary 
should 
belong to the 3D-$XY$ 
universality class \cite{Roscilde07},
and 
its end-point singularity at
$J_2 =0.5$ is our concern;
the power-law behavior of the phase boundary
(\ref{definition_of_phi})
is
characterized by
the crossover critical exponent $\phi$ \cite{Riedel69,Pfeuty74}.
In the boson language \cite{Roscilde07},
the $XY$ (paramagnetic) phase 
is interpreted as the superfluid (Mott insulator) phase.
}
\end{figure}

A schematic 
ground-state
phase diagram \cite{Dutta98}
for the square-lattice
$J_1$-$J_2$ $XY$ model
with the single-ion anisotropy
$D$
(\ref{Hamiltonian})
is presented in Fig. \ref{figure1}.
The limiting cases,
$D=0$ \cite{Bishop08}
and $J_2=0$ \cite{Pires11},
have been investigated
in depth.
As indicated, in the large-$D$ regime,
the paramagnetic phase extends irrespective of $J_2$.
In the small-$D$ regime, 
the $XY$ and
collinear phases appear, as the next-nearest-neighbor interaction $J_2$ changes.
According to the coupled-cluster-expansion method \cite{Bishop08},
the transition point was estimated as $J_2=0.50(1)$.
In the boson language \cite{Roscilde07},
the $XY$ and paramagnetic phases
correspond to the super-fluid and Mott-insulator
phases,
respectively.
The aim of this paper is to explore
the end-point singularity
(multi-criticality) of 
the phase boundary separating the $XY$ and paramagnetic phases.
For instance,
the power-law singularity of the phase boundary
\begin{equation}
\label{definition_of_phi}
D_c(J_2) -D_c(0.5) \propto (0.5-J_2)^{1/\phi}
,
\end{equation}
around $J_2(/J_1) \to 0.5^-$
is characterized by the crossover exponent
$\phi$ \cite{Riedel69,Pfeuty74}.
We determine the set of multi-critical indices
such as $\phi$, 
based on 
the
crossover-scaling theory
\cite{Riedel69,Pfeuty74}
as well as the idea of 
the quantum Lifshitz criticality
\cite{Dutta98,Hornreich75,Dutta97,Diehl00,Diehl01,Leite03,Mergulhao98,Mergulhao99}.


A notable feature of the quantum Lifshitz criticality 
\cite{Dutta98,Hornreich75,Dutta97,Diehl00,Diehl01,Leite03,Mergulhao98,Mergulhao99}
is that 
the correlation-length critical exponents,
$\nu_\perp$ and $\nu_\parallel$,
for the real-space and imaginary-time directions, respectively,
are not identical.
Because of the magnetic frustration within the real-space spin configuration,
the correlation lengths
for the real-space 
$\xi$ 
and imaginary-time 
$\xi_\tau$
directions
diverge independently as
\begin{equation}
\label{definition_of_nuperp_and_nuparallel}
\xi \sim |D-D_c|^{-\nu_\perp}
\ 
{\rm and}
\ 
\xi_\tau  (\sim \Delta E^{-1})\sim |D-D_c|^{-\nu_\parallel}
,
\end{equation}
respectively,
with the energy gap $\Delta E$
at the quantum Lifshitz point,
$J_2=0.5$.
The anisotropy is indexed 
quantitatively
by
the dynamical critical exponent
\begin{equation}
	\label{definition_of_dynamical_critical_exponent}
z = \nu_\parallel/\nu_\perp
.
\end{equation}
In the exact-diagonalization approach,
we do not have to manage the finite-size scaling
of the imaginary-time sector,
because 
the ratio
$\xi_\tau/\beta \to 0$ 
($\beta$: the inverse temperature)
vanishes
{\it a priori} at the ground state, $\beta \to \infty$;
note that the ground state is directly accessible via the exact diagonalization scheme.
Thus, 
we are able to concentrate on the finite-size scaling of the real-space sector,
and significant simplification is attained even for $z\ne 1$.

The character of
the $J_2$-driven phase transition depends on the 
magnitude of 
the
constituent spins, $S$.
As shown in Fig. \ref{figure1}, 
a direct
$XY$-collinear phase transition occurs for $S=1$ \cite{Bishop08}, whereas
an intermediate phase appears around $J_2 \approx 0.5$ as for 
the spin-half ($S=1/2$) magnet \cite{Bishop08b};
in the case of the easy-axis $XXZ$ magnet, the appearance of such an intermediate phase
remains controversial
\cite{Oitmaa20,Kalz11}.
In fact,
as mentioned above,
the kinetic frustration leads to a variety of super-solid phases
for 
the hard-core ($S=1/2$) \cite{Chen17,Tu20} and 
soft-core ($S=1$) \cite{Dong17} boson models.
Here, 
we focus our attention on the soft-core-boson case, namely, the $S=1$ $XY$ magnet \cite{Bishop08},
and show that
the end-point singularity of the super-fluid-insulator phase boundary
is under the reign of 
the above-mentioned critical exponents, $\phi$ and $\nu_{\parallel,\perp}$.

The rest of this paper is organized as follows.
In Sec. \ref{section2}, the numerical results are presented.
The finite-size scaling theory for the fidelity susceptibility is
outlined prior to presenting the simulation data.
In Sec. \ref{section3},
we present the summary and discussions.

\section{\label{section2}Numerical results}

In this section, we present the numerical results for the 
$S=1$ $J_1$-$J_2$ $XY$ magnet (\ref{Hamiltonian}).
We employed the Lanczos method for
the cluster with $N \le 5^2$ spins.
To allocate the memory space,
the symmetries possessed by the Hamiltonian (\ref{Hamiltonian}),
such as
the translational invariance, 
are incorporated
into the representation of the spin configuration
\cite{Sandwick10} 
so that the $N=5^2$ spins can be tractable
\cite{Nishiyama21,Nishiyama22}.
Thereby,
we evaluated
the fidelity susceptibility
\cite{Quan06,Zanardi06,HQZhou08,Yu09,You11}
\begin{equation}
\label{fidelity_susceptibility}
\chi_F =- \frac{1}{N} \partial_{\Delta D}^2 
 F(D,D+\Delta D)|_{\Delta D=0}
 ,
\end{equation}
with the fidelity 
$F(D,D+\Delta D)=|\langle D | D +\Delta D\rangle|$
\cite{Uhlmann76,Jozsa94,Peres84,Gorin06},
The fidelity $F$ is readily accessible via the exact diagonalization scheme,
because the ground state
$|D\rangle$ with the single-ion anisotropy $D$
is explicitly evaluated.
Actually,
the finite difference method yields the second derivative
$\partial_{\Delta D}^2 F|_{\Delta D=0}$ reliably,
because 
the leading terms,
{\it i.e.},
$F|_{\Delta D=0}=1$ and $\partial_{\Delta D} F|_{\Delta D=0}=0$,
are the constant values by definition.

To begin with,
we outline the finite-size-scaling theory for
the fidelity susceptibility $\chi_F$ (\ref{fidelity_susceptibility});
the crossover-scaling theory is
explained afterward.
The fidelity susceptibility $\chi_F$
obeys the
scaling formula
\cite{Albuquerque10}
\begin{equation}
\label{scaling_formula}
\chi_F=L^x f\left((D-D_c)L^{1/\nu}
\right)
,
\end{equation}
with the system size $L(=\sqrt{N})$,
scaling dimension $x=\alpha_F/\nu$,
critical point $D_c$, and a certain scaling function $f$.
Here, the indices, $\alpha_F$ and $\nu$,
are the fidelity-susceptibility and 
correlation-length critical exponents, respectively.
That is, the fidelity susceptibility 
and correlation length
diverge as
$\chi_F \sim |D-D_c|^{-\alpha_F}$ and 
$\xi \sim |D-D_c|^{-\nu}$, respectively.
These indices satisfy the scaling relation
\cite{Albuquerque10}
\begin{equation}
	\label{scaling_relation}
\alpha_F=\alpha+z \nu
.
\end{equation}
with the critical exponent $\alpha$
of the specific heat,
$C\sim|D-D_c|^{-\alpha}$.
Therefore,
fidelity-susceptibility's scaling dimension
$x=\alpha_F/\nu$
reduces to
\begin{equation}
\label{scaling_dimension}
	x(=\alpha_F/\nu)=2/\nu-2
,
\end{equation}
from the hyper-scaling relation
\cite{Albuquerque10}
\begin{equation}
\label{hyper_scaling_relation}
\alpha=2-(d+z)\nu
,
\end{equation}
with the real-space dimensionality $d=2$.
Notably,
the dynamical critical exponent $z$ 
cancels out in the final expression, Eq. (\ref{scaling_dimension}).
Owing to this cancellation, the $\chi_F$-mediated scaling analysis
of criticality
retrieves a substantial simplification.

\subsection{\label{section2_1}
Finite-size-scaling analysis with the fixed $J_2$:
$\chi_F$ approach}

In this section, we investigate the criticality of the 
$XY$-paramagnetic phase transition via the fidelity susceptibility $\chi_F$ 
(\ref{fidelity_susceptibility}) with 
the fixed next-nearest-neighbor interaction, $J_2= - 1/32$,   
for which a preceeding result \cite{Moura14}
is available.

\begin{figure}
	\includegraphics[width=90mm]{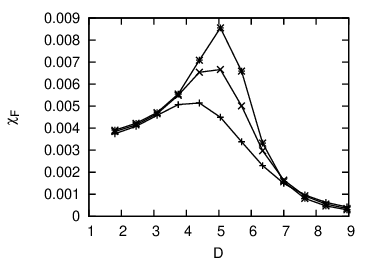}
\caption{\label{figure2}
The fidelity susceptibility
$\chi_F$ (\ref{fidelity_susceptibility})
is plotted for various
	single-ion anisotropy
	$D$ and system sizes,
($+$) $L=3$,
($\times$) $4$,
and ($*$) $5$,
with the fixed $J_2=-1/32$, 
for which a preceding self-consistent-harmonic-approximation result \cite{Moura14}
is available.
The fidelity susceptibility
exhibits a notable signature for the criticality
	at $D=D_c\approx 5$.
The critical point $D_c$
separates the $XY$ 
($D<D_c$)
and paramagnetic 
($D>D_c$)
phases.
}
\end{figure}

In Fig. \ref{figure2},
we present the fidelity susceptibility $\chi_F$ (\ref{fidelity_susceptibility})
for various single-ion anisotropy $D$ and system sizes,
($+$) $L=3$
($\times$) $4$
and
($*$) $5$,
with the fixed $J_2=-1/32$.    
The fidelity susceptibility shows a notable peak, which indicates
an onset of the $XY$-paramagnetic phase transition around $D=D_c\approx 5$.
The paramagnetic phase extends for large $D>D_c$,
whereas
the $XY$ phase is realized for small $D<D_c$ \cite{Moura14}.
In the boson language \cite{Roscilde07}, 
the single-ion anisotropy $D$ is regarded as the on-site Coulomb repulsion.
Hence, the interaction $D$ induces the Mott insulator phase from
the super-fluid phase.
Afterward, we show the energy gap, which elucidates the character of each phase
clearly.

\begin{figure}
	\includegraphics[width=90mm]{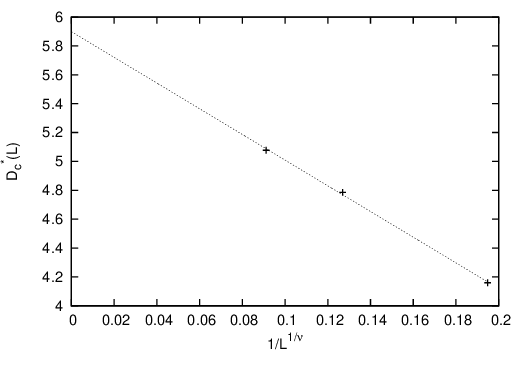}
\caption{\label{figure3}
The approximate critical point 
$D_c^*(L)$ (\ref{approximate_critical_point})
is plotted
for $1/L^{1/\nu}$ with $J_2=-1/32$. 
Here,
the correlation-length critical exponent $\nu$
is set to the value of the
3D-$XY$-universality class, $\nu=0.6717$ 
\cite{Campostrini06,Burovski06}.
The least-squares fit to these data yields an
estimate $D_c=5.90(4)$ in the thermodynamic limit
$L\to\infty$.
A possible systematic error is considered in the text.
}
\end{figure}

In order to
appreciate the critical point $D_c$
in the thermodynamic limit,
in Fig. \ref{figure3},
we present
the approximate critical point $D_c^*(L)$
for $1/L^{1/\nu}$ with the 3D-$XY$ correlation-length 
critical exponent $\nu=0.6717$
\cite{Campostrini06,Burovski06}.
Here,
the approximate critical point $D_c^*(L)$ denotes the peak position
\begin{equation}
	\label{approximate_critical_point}
\partial_D \chi_F|_{D=D_c^* (L)}=0
,
\end{equation}
of $\chi_F(L)$
for each system size $L$.
The abscissa scale $1/L^{1/\nu}$
comes from the definition of $\nu$,
{\it i.e.},
$\xi\sim |D-D_c|^{-\nu}$,
and the finite-size-scaling hypothesis, $\xi \sim L$;
therefore, the data points in Fig. \ref{figure3}  should align.
The least-squares fit to these data yields an estimate 
$D_c=5.90(4)$
for the critical point
in the thermodynamic limit $L\to\infty$.
As a reference, we also made
an alternative
extrapolation for the $L=4$-$5$ pair,
and 
arrived at 
$D_c=5.82$;
the deviation from the former one, $\approx 0.08$, 
appears to dominate the least-squares-fit error $\approx 0.04$,
Hence, regarding the deviation $0.08$ as a possible systematic error,
we estimate the critical point 
\begin{equation}
	\label{critical_point}
D_c=5.90(8)
.
\end{equation}
As a comparison,
in Table \ref{table1},
we present the preceding result \cite{Moura14}
for the two-dimensional $S=1$ $XY$ ferromagnet with the algebraically decaying
interactions,
$1/r^{10}$ ($r$: distance between spins),
by means of the self-consistent harmonic approximation (SCHA) method.
According to this study,
the critical point $D_c=6.14$ was obtained;
this value was read off from Fig. 1 of Ref. \cite{Moura14}.
Thus, the slight discrepancy between this and ours can 
be attributed
to the further neighbor ($r>\sqrt{2}$)
interactions.
Additionally, the SCHA approach is not very adequate to treat
the ground-state phase transition,
and it always overestimates the critical point, $D_c$.

\begin{table}
\caption{
The critical point $D_c$ has been estimated with
the self-consistent-harmonic-approximation
(SCHA) method \cite{Moura14}
tor
the $S=1$ $XY$ ferromagnet with the algebraically
decaying interactions,
$1/r^{10}$ ($r$: distance between spins);
the result $D_c$ was read off
from Fig.1 of Ref. \cite{Moura14}.
The present exact-diagonalization
(ED)
simulation was performed for 
the $J_1$-$J_2$ model with
$J_2= - 1/32$. 
Therefore, the slight discrepancy can be attributed to the 
further neighbor ($r> \sqrt{2}$)
interactions
	as well as the systematic deviation (overestimation)
inherent in SCHA at the ground state.
}
\label{table1}       
\begin{tabular}{lll}
\hline\noalign{\smallskip}
Method & Model & $D_c(/J_1)$ \\
\noalign{\smallskip}\hline\noalign{\smallskip}
SCHA \cite{Moura14} & algebraically decaying interactions $r^{-10}$ & $6.14$ \\
	ED (this work) & second neighbor interaction $J_2/J_1=-\frac{1}{32}$ & $5.90(8)$ \\
\noalign{\smallskip}\hline
\end{tabular}
\end{table}

\begin{figure}
	\includegraphics[width=90mm]{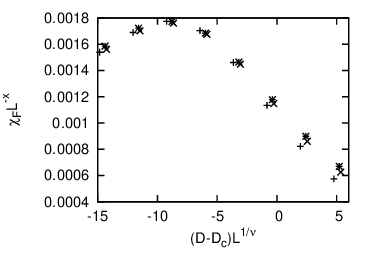}
\caption{\label{figure4}
The scaling plot,
$(D-D_c)L^{1/\nu}$ versus $L^{-x} \chi_F$,
is presented
for various $D$ and 
system sizes,
($+$) $L=3$,
($\times$) $4$, and
($*$) $5$,
with the fixed $J_2=-1/32$   
and $x=2/\nu-2$ (\ref{scaling_dimension}).
Here, the scaling parameters are set to
	$D_c=5.90$ (\ref{critical_point}),
	and $\nu=0.6717$
	for the 3D-$XY$ universality class
\cite{Campostrini06,Burovski06}.
	Hence, no {\it ad hoc} adjustable parameter is undertaken.
	The scaled data seem to fall into a scaling curve satisfactorily.
}
\end{figure}

As a cross-check,
we examine $\chi_F$'s finite-size scaling behavior, based on the
scaling formula (\ref{scaling_formula}).
In Fig. \ref{figure4},
we present the
finite-size-scaling plot, 
$(D-D_c)L^{1/\nu}$ versus $L^{-x} \chi_F$,
for various $D$ and system sizes,
($+$) $L=3$,
($\times$) $4$,
and
($*$) $5$,
with the fixed $J_2=-1/32$   
and
$x=2/\nu-2$ (\ref{scaling_dimension}).
Here, the scaling parameters are set to
$D_c=5.90$ (\ref{critical_point}) and
$\nu=0.6717$ for the 3D-$XY$ universality class
\cite{Campostrini06,Burovski06}.
We stress that no {\it ad hoc} adjustable parameter is undertaken.
The scaled data seem to fall into the scaling curve satisfactorily,
confirming the validity of the $\chi_F$-mediated analysis
of the criticality.

Last, we address a remark.
The singularity of the fidelity susceptibility, $x=\alpha_F/\nu$
(\ref{scaling_dimension}),
is larger than that of the specific heat, $\alpha/\nu$,
as the scaling relation (\ref{scaling_relation}) indicates.
Such a feature is significant in the above finite-size-scaling analyses,
because specific-heat's index $\alpha$ is negative
for the 3D-$XY$ universality class
\cite{Campostrini06,Burovski06}.
The fidelity susceptibility picks up the singularity out of
the regular part rather sensitively.
Hence, the scaling behavior is less influenced by
the finite-size artifact \cite{Yu09},
as compared to the ordinary quantifiers such as the specific heat.
Encouraged by this finding, we turn to the analysis of multi-criticality
via the probe $\chi_F$.

\subsection{\label{section2_2}
Crossover-scaling analysis around $J_2 \to 0.5^-$:
$\chi_F$ approach
}

In this section, we explore the crossover-scaling behavior 
of the fidelity susceptibility $\chi_F$ (\ref{fidelity_susceptibility})
around $J_2=0.5$
in the context of the quantum Lifshitz point
\cite{Dutta98};
more specifically,
we investigate 
how the phase boundary ends up at $J_2=0.5$.
For that purpose, 
we have to extend the scaling formula
(\ref{scaling_formula}) so as to include
yet another scaling parameter, 
$0.5-J_2$ (distance from the multi-critical point),
and an accompanying crossover exponent $\phi$.
According to the crossover-scaling theory 
\cite{Riedel69,Pfeuty74},
the extended scaling formula should read
\begin{equation}
\label{crossover_scaling_formula}
\chi_F = L^{\dot{x}}
g
\left(
(D-D_c(J_2))L^{1/\nu_\perp} ,  (0.5-J_2)L^{\phi/\nu_\perp} 
\right)
,
\end{equation}
with 
the multi-critical scaling
dimension
$\dot{x}$,
crossover exponent $\phi$,
distance from the multi-critical point, $0.5-J_2$,
and a certain scaling function $g$.
As in Eq. (\ref{scaling_dimension}),
the multi-critical scaling dimension 
$\dot{x}$ is given by
\begin{equation}
\label{multi-critical_scaling_dimension}
	\dot{x}(=
	\dot{\alpha}_F/\nu_\perp)=2/\nu_\perp -2
,
\end{equation}
with the multi-critical fidelity susceptibility 
critical exponent $\dot{\alpha}_F$,
{\it i.e.},
$\chi_F \sim |D-D_c|^{-\dot{\alpha}_F}$
and
the multi-critical correlation length critical exponent
$\nu_\perp$ 
(\ref{definition_of_nuperp_and_nuparallel}).
Notably, the scaling dimension $\dot{x}=2/\nu_\perp-2$
(\ref{multi-critical_scaling_dimension}) is not
affected by $z$.
The exponent $z$ is thus considered in the next section.
As mentioned in Introduction,
the crossover exponent $\phi$ is relevant to
the algebraic singularity of the phase boundary,
$D_c \sim (0.5-J_2)^{1/\phi}$ (\ref{definition_of_phi}).
This relation
immediately follows from the above scaling formula (\ref{crossover_scaling_formula}):
Because
the second argument of the scaling formula (\ref{crossover_scaling_formula})
is scale-invariant (dimensionless),
the relation
$(0.5-J_2)^{1/\phi}\sim L^{-1/\nu_\perp}$ holds.
The definition of $\nu_\perp$,
{\it i.e.},
$\xi \sim |D-D_c|^{- \nu_\perp}$
(\ref{definition_of_nuperp_and_nuparallel}), 
and the scaling hypothesis, $L \sim \xi$,
admit
the desired expression,
$D_c \sim (0.5-J_2)^{1/\phi}$ (\ref{definition_of_phi}).

\begin{figure}
	\includegraphics[width=90mm]{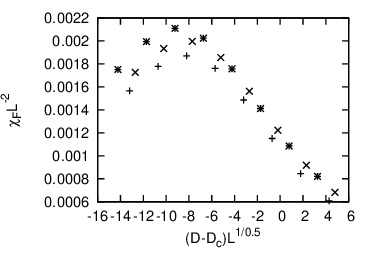}
\caption{ \label{figure5}
The crossover-scaling plot,
$(D-D_c(J_2)) L^{1/\nu_\perp}$ versus $L^{-\dot{x}} \chi_F$,
is presented 
for various $D$ and
system sizes,
($+$) $L=3$,
($\times$) $4$, and
($*$) $5$,
with the fixed
crossover exponent
$\phi=0.6$,
multi-critical correlation-length exponent
$\nu_\perp=0.5$,
and
multi-critical scaling dimension
	$\dot{x}(=2/\nu_\perp-2 )=2 $ (\ref{multi-critical_scaling_dimension}).
The second argument of the crossover-scaling formula (\ref{crossover_scaling_formula_gap})
is fixed to $(0.5-J_2)L^{\phi/\nu_\perp}=0.72$.
	The $L=4$ ($\times$)
	and $5$ ($*$)
	data
	are almost overlapping each other.
}
\end{figure}

We then apply the scaling theory (\ref{crossover_scaling_formula})
to the analysis of the simulation data.
In Fig. \ref{figure5},
we present the crossover-scaling plot,
$(D-D_c(J_2))L^{1 / \nu_\perp}$ versus $L^{-\dot{x}}\chi_F$,
for various $D$ and
system sizes,
($+$) $L=3$,
($\times$) $4$,
and
($*$) $5$,
with the 
multi-critical
scaling dimension 
$\dot{x}=2/\nu_\perp-2$
(\ref{multi-critical_scaling_dimension}), and
the
critical point $D_c(J_2)$
determined via the same scheme as that of Sec. \ref{section2_1}.
Here,
the second argument of the crossover-scaling formula 
(\ref{crossover_scaling_formula})
fixed to $(J_2-0.5)L^{\phi/\nu_\perp}=0.72$
with the 
optimal values of the multi-critical indices,
$\phi=0.6$ and $\nu_\perp=0.5$.
In Fig. \ref{figure5},
we see that the crossover-scaled data collapse into a scaling curve.
Actually,
around the critical point,
$(D-D_c)L^{1/\nu_\perp} \approx 0$ (right-side slope),
in particular,
the data for $L=4$ ($\times$)
and $5$ ($*$) are almost overlapping each other. 
Likewise, we made similar crossover-scaling analyses for other values of
$\phi$ and $\nu_\perp$,
and found that these indices lie within
\begin{equation}
\label{phi}
\phi = 0.60(15),
\end{equation}
and
\begin{equation}
\label{nuperp}
\nu_\perp=0.5(1)
,
\end{equation}
respectively.

We address a few remarks.
First, 
the underlying mechanisms behind the scaling plots, Fig. \ref{figure4}
and \ref{figure5}, are not identical.
Actually,
the
scaling dimensions 
of the former, 
$x\approx 0.98\dots$ (\ref{scaling_dimension}),
and the latter,
$\dot{x}=2$ 
(\ref{multi-critical_scaling_dimension}),
are far apart.
Therefore,
the data collapse of the crossover-scaling plot in
Fig. \ref{figure5}
is by no means accidental;
actually,
both $\phi$ and $\nu_\perp$ have to be tuned carefully.
Last,
the direct numerical simulation right at $J_2=0.5$ suffers from finite-size artifact
due to
the $J_2$-mediated precursor to the super-lattice structures (stripe patterns),
which may conflict with the system size $L$.
The crossover-scaling formula (\ref{crossover_scaling_formula})
provides an alternative route toward $J_2 \to 0.5^-$
to attain smooth finite-size-scaling behaviors.

\subsection{\label{section2_3}
Crossover-scaling analysis around $J_2 \to 0.5^-$:
$\Delta E$ approach
}

In this section, we estimate the dynamical critical exponent $z$ 
(\ref{definition_of_dynamical_critical_exponent})
at the quantum Lifshitz point.
For that purpose, we consider the energy gap $\Delta E$.
The energy gap should obey
the same crossover-scaling formula
\cite{Riedel69,Pfeuty74}
as that of $\chi_F$ (\ref{crossover_scaling_formula})
\begin{equation}
\label{crossover_scaling_formula_gap}
\Delta E = L^{\dot{y}}
h
\left(
(D-D_c(J_2))L^{1/\nu_\perp} ,  (0.5-J_2)L^{\phi/\nu_\perp} 
\right)
,
\end{equation}
with 
$\Delta E$'s multi-critical scaling dimension $\dot{y}$,
and
 a certain scaling function $h$.
The
scaling dimension 
$\dot{y}$
is given by 
the dynamical critical exponent
\begin{equation}
\label{scaling_formula_gap}
\dot{y}=    -   z
,
\end{equation}
because of Eq.
(\ref{definition_of_nuperp_and_nuparallel}),
(\ref{definition_of_dynamical_critical_exponent}) and the scaling hypothesis, $\xi\sim L$.
Technically, the energy gap was evaluated by the expression
\begin{equation}
	\label{definition_of_energy_gap}
\Delta E = E_0(1)-E_0(0)
.
\end{equation}
Here, the symbol $E_0 (S^z_{tot})$
denotes
the ground state energy within the Hilbert-space sector
indexed by the total longitudinal magnetic moment $S^z_{tot}$.
In the language boson,
this quantity measures the energy cost due to 
the increment by one particle.

\begin{figure}
	\includegraphics[width=90mm]{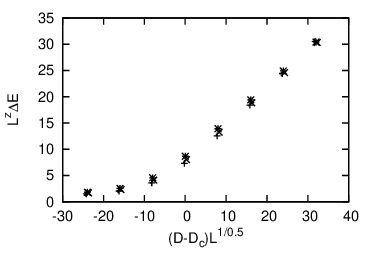}
\caption{\label{figure6}
The crossover-scaling plot,
$(D-D_c(J_2)) L^{1/\nu_\perp}$ versus $L^z\Delta E$,
is presented 
for various
$D$ and
system sizes,
($+$) $L=3$,
($\times$) $4$, and
($*$) $5$,
	with an optimal dynamical critical exponent
$z=1.8$. 
The second argument of the crossover-scaling formula (\ref{crossover_scaling_formula_gap})
is fixed to $(0.5-J_2)L^{\phi/\nu_\perp}=0.72$.
The indices $\nu_\perp=0.5$ and $\phi=0.6$
are the same as those of Fig. \ref{figure5}.
}
\end{figure}

In Fig. \ref{figure6},
we present the crossover-scaling plot, 
$(D-D_c(J_2)) L^{1/\nu_\perp}$ versus $L^z \Delta E$,
for various $D$ and 
system sizes,
($+$) $L=3$,
($\times$) $4$,
and
($*$) $5$,
with an optimal value of the dynamical critical exponent $z=1.8$.
Here,
the second argument of the crossover-scaling formula
(\ref{crossover_scaling_formula_gap}) is fixed to
$(0.5-J_2)L^{\phi/\nu_\perp}=0.72$,
as in Fig. \ref{figure5}.
The scaling parameters,
$\nu_\perp=0.5$ and
$\phi=0.6$, are also taken from 
Fig. \ref{figure5}.
In Fig. \ref{figure6},
the scaled data seem to collapse into a scaling curve satisfactorily.
Such a feature confirm the validity of the crossover-scaling analysis
of $\chi_F$ in Fig. \ref{figure5}.
We made similar data analyses for other values of $z$,
and conclude that 
the dynamical critical exponent lies within
\begin{equation}
\label{dynamical_critical_exponent}
z=1.8(2)
.
\end{equation}
Note that for the ordinary phase transition including the 3D-$XY$ universality,
the dynamical critical exponent should be $z=1$.
Therefore, the value $z=1.8$ indicates that the multi-criticality at $J_2=0.5$
is characterized by the anisotropy between the real-time and imaginary-time subspaces.
The dynamical critical exponent $z=1.8(2)$ 
[Eq. (\ref{dynamical_critical_exponent})] determines
the correlation-length critical exponent 
$\nu_\parallel$ along
the imaginary-time direction
\begin{equation}
\label{nuparallel}
\nu_\parallel=0.9(2)
,
\end{equation}
through the relations, Eq.
(\ref{definition_of_nuperp_and_nuparallel})
and
(\ref{definition_of_dynamical_critical_exponent}).

This is a good position to make a comparison between
the preceeding field-theoretical result \cite{Diehl01} and ours.
As mentioned in Introduction,
the end-point singularity of the $XY$-paramagnetic phase boundary
(multi-criticality) has been investigated in the context of the quantum Lifshitz point
\cite{Dutta98}.
The $\varepsilon$-expansion results up to O$(\varepsilon^2)$ are reported in Ref. \cite{Diehl01};
in these expressions,
the parameters, frustrated subspace dimension $m$ and 
number of order-parameter's components $N$,
have to be set to $m=2$ and $N=2$, respectively,
and
the expansion parameter $\varepsilon$ has to be set to $\varepsilon=2$
because of the upper critical dimension $d^*=5$ \cite{Diehl01}.
In Table \ref{table2}, the raw $\varepsilon$-expansion results as well as the 
$[1:1]$
Pad\'e approximants
for the indices, $\nu_\perp$, $\nu_\parallel$, and $\phi$, are shown.
These two results look alike except $\nu_\parallel$;
the deviations between them may be regarded as indicators of uncertainty.
Within the error margins,
our results agree with those of the $\varepsilon$-expansion method.
Specifically, as to $\phi$,
our result supports the raw data.
Hence, the end-point singularity
for the $S=1$ $J_1$-$J_2$
$XY$ model (\ref{Hamiltonian})  can be
identified as
the quantum Lifshitz critical point \cite{Dutta98}.

\begin{table}
\caption{
The two-dimensional O$(2)$
quantum Lifshitz criticality 
\cite{Dutta98}
has been
investigated with the $\varepsilon$-expansion ($\varepsilon$ exp.)
method up to O$(\varepsilon^2)$ \cite{Diehl01}.
The results for
the multi-critical indices, $\nu_\perp$, 
$\nu_\parallel(=z\nu_\perp)$,
and $\phi$,
as well as the $[1:1]$ Pad\'e approximants
are shown.
The expansion parameter has to be set to $\varepsilon=2$ because of
	the upper critical dimension $d^*=5$ \cite{Diehl01}.
}
\label{table2}       
\begin{tabular}{llll}
\hline\noalign{\smallskip}
indices & O$(\varepsilon^2)$ $\varepsilon$ exp. \cite{Diehl01} & 
$[1:1]$ Pad\'e & this work \\
\noalign{\smallskip}\hline\noalign{\smallskip}
	$\nu_\perp$ & $0.421$ & $0.600$ & $0.5(1)$ \\
	$\nu_\parallel$ & $0.867$ & $1.696$ & $0.9(2)$ \\
	$\phi$ & $0.661$ & $0.763$ &   $0.60(15)$ \\
\noalign{\smallskip}\hline
\end{tabular}
\end{table}

Last, we address a remark.
The $\Delta E$ data in Fig. \ref{figure6} show that
in the paramagnetic ($XY$) phase, $D>(<)D_c$,
the energy gap opens (closes);
note that the ordinate axis is scaled by $L^z$.
Actually,
as the formula
(\ref{definition_of_nuperp_and_nuparallel}) indicates,
the gap opens  almost linearly as
\begin{equation}
	\Delta E (\sim |D-D_c|^{\nu_\parallel})
	\sim |D-D_c|^{0.9}
	,
\end{equation}
in the $D>D_c$ side.
Hence, the energy gap captures the character of each phase clearly.
In contrast, as shown in Fig. \ref{figure5},
the fidelity susceptibility is rather sensitive
to the onset of the phase transition.
The gapless excitation in the $XY$ phase corresponds to the Goldstone mode
(spin wave), whereas
the finite magnon mass indicates the paramagnetism.
In the boson language, this finite magnon mass 
is interpreted as the Mott-insulator gap induced by the on-site Coulomb repulsion
$D$;
see the formula (\ref{definition_of_energy_gap}) for $\Delta E$.

\section{\label{section3}
Summary and discussions}

The $S=1$ square-lattice $J_1$-$J_2$ $XY$ magnet 
with the single-ion anisotropy
$D$
(\ref{Hamiltonian})
was investigated with an emphasis on the end-point singularity of the
$XY$-paramagnetic phase boundary toward $J_2\to0.5^-$.
To surmount 
the negative sign problem of the QMC method, we employed the exact diagonalization method,
and
as a probe to detect the criticality,
we implemented the fidelity susceptibility $\chi_F$ (\ref{fidelity_susceptibility}).
A benefit of the $\chi_F$-mediated approach is that the scaling dimension (\ref{scaling_dimension})
is not affected by
the dynamical critical exponent $z(\ne 1)$.
As a demonstration,
under the setting
$J_2=-1/32$ \cite{Moura14},
the $D$-driven phase transition was analyzed via $\chi_F$.
Based on the scaling formula (\ref{scaling_formula}),
we confirmed that the 
$XY$-paramagnetic phase transition
belongs to the
3D-$XY$ universality class rather clearly.
Thereby,
the $\chi_F$ data were cast into the 
crossover scaling formula (\ref{crossover_scaling_formula})
with the properly scaled
distance from the multi-criticality,
$0.5-J_2\sim L^{-\phi/\nu_\perp}$.
As a result,
the crossover exponent $\phi$
and the multi-critical correlation-length exponent
$\nu_\perp$
are estimated
as $\phi=0.60(15)$ 
[Eq. (\ref{phi})]
and $\nu_\perp=0.5(1)$
[Eq. (\ref{nuperp})], respectively.
Additionally, making the similar crossover-scaling analysis of the
energy gap $\Delta E$,
we obtained
the dynamical critical exponent $z=1.8(2)$
[Eq. (\ref{dynamical_critical_exponent})].
As shown in Table \ref{table2},
these critical indices agree with those of the
quantum Lifshitz criticality
\cite{Dutta98,Diehl01}
within the error margins of both analyses.

Our result shows that the 
$XY$-paramagnetic phase boundary terminates
at
$J_2 \to 0.5^-$, obeying the quantum-Lifshitz-point scenario.
For the Ising counterpart, the similar conclusion was
obtained with the extensive series-expansion analysis \cite{Oitmaa20}.
As for the $XXZ$ case,
intermediate phases are induced
around the multi-critical point \cite{Kalz11}.
Therefore, it would be tempting to consider
how the above quantum Lifshitz criticality changes,
as the extended interactions are turned on.
This problem is left for the future study.

\section*{Acknowledgment}
This work was supported by a Grant-in-Aid
for Scientific Research (C)
from Japan Society for the Promotion of Science
(Grant No. 
20K03767).

	\section*{Data Availability Statement}

	My manuscript has no associated data.

Data will be made available on reasonable request.



%

%
%




\end{document}